\documentclass[journal,onecolumn]{IEEEtran}
\usepackage{epsfig}
\usepackage{graphicx}
\usepackage{graphics}
\usepackage{epstopdf} 
\usepackage{tikz}
\usetikzlibrary{calc,shapes.geometric}
\usetikzlibrary{arrows,snakes,backgrounds}

\ifCLASSINFOpdf
\else
\fi

\usepackage[cmex10]{amsmath}

\def\clock{\n=\time \divide\n 60
  \m=-\n \multiply\m 60 \advance\m \time
  \ifnum \n>12 \advance\n -12 \fi
   \number\n.\twodigits\m~\ampm\time}
\def\ampm#1{\ifnum #1< 720 am\else pm\fi}
\def\twodigits#1{\ifnum #1<10 0\fi \number#1}

\usepackage{epsfig}

\def\hyptest{\renewcommand{\arraystretch}{-0.7} 
\begin{array}{c}  
\mbox{\tiny{$H_{1}$}}  \\ \vspace{-0.5 mm}
>\\ 
<\\  
\mbox{\tiny{$H_{0}$}} 
\end{array}
}

\def\nexto{\kern -0.54em}

\def\prob{{\rm {I\ \nexto P}}}

\def\pfa{{\rm P_{FA}}}
\def\pd{{\rm P_{D}}}

\newcount\m \newcount\n
\def\clock{\n=\time \divide\n 60
  \m=-\n \multiply\m 60 \advance\m \time
  \ifnum \n>12 \advance\n -12 \fi
   \number\n.\twodigits\m~\ampm\time}
\def\ampm#1{\ifnum #1< 720 am\else pm\fi}
\def\twodigits#1{\ifnum #1<10 0\fi \number#1}

\begin{document}

\title{An Introduction to Radar Sliding Window Detectors}
\author{G. V. Weinberg  \\National Security and Intelligence, Surveillance and Reconnaissance Division\\Defence Science and Technology Group\\
P. O. Box 1500 Edinburgh, South Australia 5111, Australia \\ (Draft created at \clock)\\
Graham.Weinberg@dst.defence.gov.au
 }
\maketitle

\markboth{Introduction to Radar Sliding Window Detectors,  \today}%
{}

\begin{abstract}
An introduction to the theory of sliding window detection processes, used as alternatives to optimal Neyman-Pearson based radar detectors, is presented. Included is an outline of their historical development, together with an explanation for  the resurgence of interest in such detectors for operation in modern maritime surveillance radar clutter. In particular, recent research has developed criteria that enables one to construct such detection processes with the desired constant false alarm rate property for a comprehensive class of clutter model. The chapter also includes some examples of the performance of such detectors in a specific interference environment.
\end{abstract}


\section{Introduction}
Sliding window detectors were first introduced in the 1960s and investigated because such detectors could achieve certain desirable features not easily acquired with decision rules based upon the Neyman-Pearson Lemma.
During the 1960s most radar systems were of low spatial resolution (relative to today) and for these radar systems it was found that the backscatter from surface clutter, such as the sea surface, could be represented as a Gaussian process. 
In particular, in terms of amplitude statistics, the Rayleigh assumption was valid, which in amplitude squared corresponded to exponentially distributed clutter.
Based upon such a clutter model assumption the pioneering work of \cite{finn} demonstrated that a suboptimal detection scheme could be specified, which achieved what is known as a constant false alarm rate (CFAR). This means that an adaptive test could be proposed, 
where a cell under test (CUT) is compared with an appropriate measurement statistic of the clutter such that when the measurement statistic was appropriately normalised, the resultant detection threshold produced an (ideally) fixed probability of false alarm. This resulted in a loss relative to an optimal decision rule, but the scheme's implementation was simpler and did not require an {\em a-priori} estimate of the radar target's strength.

As has been demonstrated in \cite{levanonbook}, the problem with optimal detectors is that if an estimate of a model parameter is applied to a detection scheme, it can result in unacceptable variation in the false alarm rate. An increase in the number of false alarms can result in missed detection of real targets, operator confusion and hence can be unacceptable in an operational radar system. Hence sacrificing some detection performance (i.e. a sufficiently small and predictable CFAR loss) has been considered acceptable if the CFAR property can be achieved. Consequently much effort was focused on the development of sliding window detectors, for operation in exponentially distributed clutter, especially since it was found that many such detectors could be proposed. Since all of these detectors need to make an estimate of a statistic of the clutter in the vicinity of the CUT, and clutter is often anisotropic and memory was expensive, the practical implementation was to use a sliding window estimate of the statistic computed from data near to the CUT. See for example, the works of \cite{nitzberg} - \cite{gandhi88}.

Figure \ref{fig1} provides an illustration of the mechanics of a sliding window detection process. A series of measurements are assumed available, from which a set of clutter statistics is extracted, denoted $C_j$ in the figure. This set is referred to as the clutter range profile (CRP). These are separated from the CUT by a number of guard cells, whose purpose is to limit the effects of a range spread target. These clutter measurements are then compressed to two measurements denoted by $p_1$ and $p_2$, which are then combined to produce a single measurement of the clutter level ($g$). The latter is then normalised by $\tau > 0$ and
a comparison is then made to determine the presence of a target in the CUT. The next range bin is then (conceptually) tested in the same manner by moving all test cells down one range bin. The threshold crossing results from this process are then passed to subsequent stages of the signal and data processing (which may include combining data from multiple pulses, automatic tracking system, and displays).

\tikzstyle{normalisation}=[circle,draw=blue!50,fill=blue!20,thick,
inner sep=0pt,minimum size=18mm]
\tikzstyle{testcellA}=[rectangle,draw=black!50,fill=green!20,thick,
inner sep=0pt,minimum size=10mm]
\tikzstyle{processX}=[rectangle,draw=blue!40,fill=blue!20,thick,
inner sep=0pt, minimum size=18mm]
\tikzstyle{processY}=[rectangle,draw=black!50,fill=red!20,thick,
inner sep=0pt, minimum size=15mm]
\tikzstyle{processZ}=[rectangle,draw=black!50,fill=green!30,thick,
inner sep=0pt, minimum size=15mm]
\tikzstyle{decisionT}=[
   isosceles triangle,
   draw=black!50,fill=black!10,
   shape border rotate=0,
   inner sep=3,
   isosceles triangle apex angle=60,
   isosceles triangle stretches,
   minimum size = 15mm]

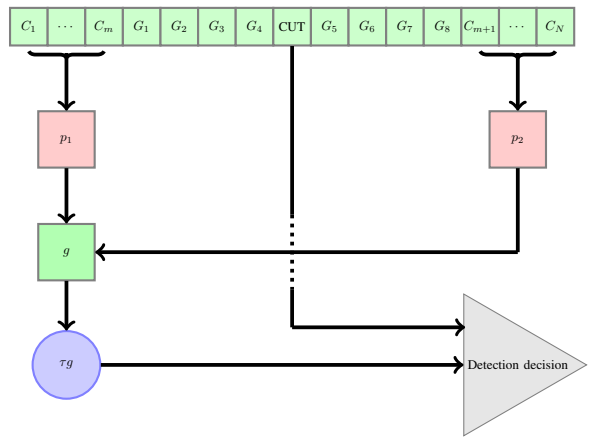
\begin{figure}
\centering
\begin{tikzpicture}[scale =0.5, transform shape]

\node (input) at (-2, 0) [testcellA] {$C_1$};
\node at (-1, 0) [testcellA] {$\cdots$};
\node (inputend) at (0, 0) [testcellA] {$C_m$};
\node at (1,0) [testcellA] {$G_1$};
\node at (2,0) [testcellA] {$G_2$};
\node at (3, 0) [testcellA] {$G_3$};
\node at (4, 0) [testcellA] {$G_4$};
\node (cut) at (5,0) [testcellA] {CUT};
\node at (6, 0) [testcellA] {$G_5$};
\node at (7,0) [testcellA] {$G_6$};
\node at (8, 0) [testcellA] {$G_7$};
\node at (9, 0) [testcellA] {$G_8$};
\node (inputEA) at (10, 0) [testcellA] {$C_{m+1}$};
\node at (11,0) [testcellA] {$\cdots$};
\node (inputEB) at (12, 0) [testcellA] {$C_{N}$};
\node (processA) at (-1, -3) [processY] {$p_1$};
\node (processB) at (11, -3) [processY] {$p_2$};
\node (processC) at (-1, -6) [processZ] {$g$};
\node (processD) at (-1, -9) [normalisation] {$\tau g$};
\node (decision) at (11, -9) [decisionT] {Detection decision};

\draw [->, line width=0.5mm] (-1,-0.7) -> (processA.north);
\draw [->, line width=0.5mm] (11,-0.7) -> (processB.north);

\draw [-, line width=0.5mm] (processB.south) to  (11, -6);
\draw [->,  line width=0.5mm] (11, -6) to  (processC.east);
\draw [-, line width=0.5mm]  (cut.south) to (5, -5);
\draw [-, dotted, line width=0.5mm]  (5, -5) to (5, -7);
\draw [-, line width=0.5mm]  (5,-7) to (5, -8);
\draw [->, line width=0.5mm]  (5,-8) to (9.6, -8);

\draw [->, line  width=0.5mm] (processA.south) -- (processC.north);
\draw [->, line  width=0.5mm] (processC.south) -- (processD.north);
\draw [->, line  width=0.5mm] (processD.east) -- (decision.west);

\draw[snake=brace,mirror snake, black, line width=0.5mm, raise snake=2pt] (input.south) -- (inputend.south);
\draw[snake=brace,mirror snake, black, line width=0.5mm, raise snake=2pt] (inputEA.south) -- (inputEB.south);

\end{tikzpicture}

\caption{An illustration of a sliding window detection process. Clutter measurements ($C_j$) are used to produce a normalised measurement of the clutter level. This is then compared with the cell under test  (CUT) and a decision on the presence of a target can be made. The guard cells (arbitrary 8 in this example) are denoted $G_1$ to $G_8$.}\label{fig1}
\end{figure}

Although detectors such as that illustrated in Figure \ref{fig1} could be formulated with the CFAR property in exponentially distributed clutter, as radar resolution was increased (finer range cells, narrower beams), the clutter statistics were found to change radically such that the performance of these detectors either failed to achieve the CFAR property and/or produced an unacceptably large CFAR loss. As an example we will consider maritime surveillance radar systems where changes to resolution and/or viewing geometry can produce radical changes to the observed clutter statistical distributions.
In the case where clutter can be modelled by log normal or Weibull distributions \cite{goldstein} identified such a process with the full CFAR property. A similar and somewhat simpler detector was reported in \cite{haykin}, based upon order statistics.
Other studies focused on various ways in which the CFAR property could be achieved; see for example \cite{anast95} and \cite{watts}. However, in high resolution radar, the sea clutter has been found to be modelled appropriately by the K-distribution \cite{watts}  and for this distribution the approaches employed for designing low resolution  sliding window CFAR detectors failed to achieve the CFAR property and produced large CFAR losses.

Observing that our primary interest in representing clutter as a statistical distribution is in the "long tail", the Pareto family of distributions has been proposed as a more tractable alternative to the K-distribution \cite{balleri} - \cite{weinberg11}. This distribution is simpler than the K-distributional model but fits the currently accepted framework for radar clutter models, since the Pareto Type II
distribution arises as an intensity model of a compound Gaussian process with inverse gamma texture \cite{balleri}.  Based upon this new distribution it became possible to produce sliding window detectors with the CFAR property, as in the exponentially distributed clutter case.
The first such contribution is \cite{weinberg13a}, who showed that a transfromation approach could be developed to produce sliding window detectors for operation in Pareto Type I clutter, such that they achieved the CFAR property with respect to the Pareto shape parameter, but required {\em a priori} knowledge of the Pareto scale parameter. The assumption of a Pareto Type I model was justified because estimates for the Pareto scale parameter, based upon Defence Science and Technology (DST) Group's real data, showed that the scale parameter estimate tended to be very small, and so the Pareto Type I model could be used as an approximation for the Pareto Type II distribution \cite{weinberg11}.

This transformation approach was extended in \cite{weinberg14} and shown to be more general than the original development in \cite{weinberg13a}. Further studies of CFAR sliding window detection processes have been documented in \cite{weinberg14b} - \cite{weinbergglenny17}.
An interesting discovery was subsequently reported in \cite{weinberg17a}, where it was shown that the transformed detectors studied in \cite{weinberg13a} could be modified to achieve the full CFAR property in Pareto Type I clutter. The key to this was to replace the Pareto scale parameter in transformed detectors with a minimum of the statistics used for the clutter measurements.  Further examination of this revealed that it is actually possible to identify a class of clutter models and a generic decision rule which is CFAR. The class of distributions must be scale and power invariant and is documented in \cite{weinberg17b}. This discovery has allowed the specification of a large number of decision rules for operation in Weibull, log normal and Pareto Type I clutter which are completely CFAR.

Although it is now possible to achieve the full CFAR property in Pareto Type I clutter, there has been some interest in attempting to do this directly to the case of clutter modelled by a Pareto Type II distribution. Recently \cite{weinberg17c} examined this and showed that CFAR could be achieved with repect to the shape parameter but not the scale. This has opened up many future research directions at DST Group.

The next section formulates the sliding window detector problem mathematically and formulates some classical detectors for operation in exponentially distributed clutter.

\section{Mathematical Formulation}
The sliding window detector illustrated in Figure \ref{fig1} is now formulated mathematically. A classic reference on such detectors is \cite{minkler90} while \cite{weinbergbook} provides a modern view from the Pareto clutter model perspective. 
Suppose that the statistic of the CUT is $Z_0$ and that the CRP is modelled by the statistics $Z_1, Z_2, \ldots, Z_N$. It is assumed that all these statistics are independent and that they have the same common distribution function, including $Z_0$ in the absence of a target in the CUT.
Let $H_0$ to be the hypothesis that the CUT does not contain a target, and $H_1$ the alternative hypothesis that the CUT contains a target embedded within clutter. Then the binary test can be specified in the form 
\begin{equation}
Z_0 \hyptest \tau g(Z_1, Z_2, \ldots, Z_N), \label{primarytest}
\end{equation}
where the notation employed in \eqref{primarytest} means that $H_0$ is rejected in the case where $Z_0$ exceeds $\tau g(Z_1, Z_2, \ldots, Z_N)$.
The threshold multiplier $\tau$ is used so that the detection process \eqref{primarytest} can have its Pfa controlled adaptively.  The corresponding Pfa is given by the expression
\begin{equation}
\pfa = \prob(Z_0 > \tau g(Z_1, Z_2, \ldots, Z_N)| H_0), \label{primarytestpfa}
\end{equation}
where $\prob$ denotes probability. If the expression \eqref{primarytestpfa} is such that the threshold multiplier $\tau$ can be set independently of the clutter power, then the test \eqref{primarytest} will be referred to as a sliding window detector with the CFAR property. Since clutter power is a function of the clutter model's parameters, it is sufficient to show that $\tau$ does not depend on unknown clutter parameters to ascertain that \eqref{primarytest} is CFAR.
The corresponding probability of detection (Pd) is given by
\begin{equation}
\pd = \prob(Z_0 > \tau g(Z_1, Z_2, \ldots, Z_N)| H_1), \label{primarytestpd}
\end{equation}
and requires knowledge of the target model in order to extract a detection probability.

In the case where the clutter statistics have an exponential distribution with parameter $\lambda>0$ and distribution function
\begin{equation}
\prob(Z_j \leq t) = 1 - e^{-\lambda t}
\end{equation}
for each $j \in \{1, 2, \ldots, N\}$ and for $Z_0$ under $H_0$, where $t \geq 0$, the test \eqref{primarytest} will have the CFAR property provided $g$ is a scale-invariant function. This is the requirement that for all $\eta > 0$ 
\begin{equation}
g(\eta Z_1, \ldots, \eta Z_N) = \eta g(Z_1, \ldots, Z_N).
\end{equation}
Examples of scale-invariant functions include sums, order statistics and geometric means.

To examine some specific detectors, consider the case of target detection in exponentially distributed clutter, where the target model is Gaussian in the complex domain, or exponential in intensity.  
Then based upon the formulation in \cite{gandhi88}, this test can be specified by
$H_0: \mu = \lambda$ against the alternative $H_1: \mu = \frac{\lambda}{1+S}$, where $\mu$ represents the distributional reciprocal mean and $S$ is the signal to clutter ratio (SCR), and $\lambda$ is the clutter parameter.
Then selecting $g$ to be a sum of its arguments, the detector 
\begin{equation}
Z_0 \hyptest \tau \sum_{j=1}^N Z_j
\label{appBeq1}
\end{equation}
is known as the cell-averaging (CA) CFAR, and it can be shown that its
probability of detection is given by
\begin{eqnarray}
\pd 
&=& \left[1 + \frac{\tau}{1+S}\right]^{-N}. \label{appBeq2}
\end{eqnarray}
To extract the expression for the Pfa, one sets $S=0$ in (\ref{appBeq2}) to yield
\begin{equation}
\pfa = (1+\tau)^{-N}, \label{appBeq3}
\end{equation}  
confirming that the sliding window detector based upon \eqref{appBeq1} is indeed CFAR. It has been shown that this CA-CFAR is the optimal sliding window CFAR, for the hypothesis test under consideration \cite{gandhiadd}. In practical application of \eqref{appBeq1} one extracts $\tau$, for a given Pfa, via \eqref{appBeq3}.

A linear threshold detector, for this detection scenario, is given by
\begin{equation}
Z_0\hyptest - \frac{1}{\lambda} \log(\pfa),
\label{appBaux4}
\end{equation}
which is often used as an upper bound on performance. This is not a CFAR process since it requires {\em a priori} knowledge of the exponential shape parameter. The way in which such a detector is motivated is that if one has perfect
knowledge of the clutter parameter, then there is no need to apply a scale-invariant function to estimate it. Hence a fixed threshold detector should provide one with ideal performance \cite{anast95}. In addition to this it can be shown that the CA-CFAR's probability of detection limits to that of \eqref{appBaux4}
as $N \rightarrow \infty$ \cite{weinbergbook}. 

One of the issues with the CA-CFAR \eqref{appBeq1} is its inability to deal with interference adequately, as will be shown in the next section. Thus \cite{rohling83} introduced the ideal of using an order statistic (OS) to estimate the clutter level.
The  OS-CFAR is given by
\begin{equation}
Z_0 \hyptest \tau Z_{(k)},
\label{appBdet2}
\end{equation}
operating in exponentially distributed clutter with $1 \leq k \leq N$ and under the assumption of a Gaussian target model as before. 
Selection of the index $k$ can be based upon the desire to manage an expected numer of interfering targets.
The probability of detection corresponding to \eqref{appBdet2} is given by
\begin{eqnarray}
\pd 
&=& \frac{N!}{(N-k)!}\frac{ \Gamma(  (1+S)^{-1} \tau + N - k+1)}{\Gamma( (1+S)^{-1} \tau + N + 1)}, \label{appBeq4}
\end{eqnarray}
where $\Gamma$ is the gamma function.
Setting $S=0$ in (\ref{appBeq4}) yields the false alarm probability
\begin{equation}
\pfa = \frac{N!}{(N-k)!}\frac{ \Gamma(   \tau + N - k+1)}{\Gamma(  \tau + N + 1)}, \label{appBeq5}
\end{equation}
which is applicable regardless of the underlying target model. It is necessary to apply numerical inversion to extract $\tau$ from \eqref{appBeq5} for application in \eqref{appBdet2}.

The two detection processes \eqref{appBeq1} and \eqref{appBdet2} are two classic examples of  CFAR sliding window detectors for exponentially distributed clutter. In the next section some examples of their performance are included.

\section{Performance Examples}
In the following the length of the CRP is set to $N = 32$ while the Pfa is set to $10^{-4}$. In all situations a Gaussian target model has been applied as in the previous section's formulation. Interfering targets are generated as independent Gaussian target modes as for the CUT.
Detection performance in the absence of interference is provided by \eqref{appBeq2} and \eqref{appBeq4}. In all other cases Monte Carlo simulations are used to estimate the Pd, with $10^6$ runs for each target SCR in the CUT.

\begin{figure}[ht]
\centering
\begin{tabular}{c}
\includegraphics[width=0.5\textwidth,height=0.4\textwidth]{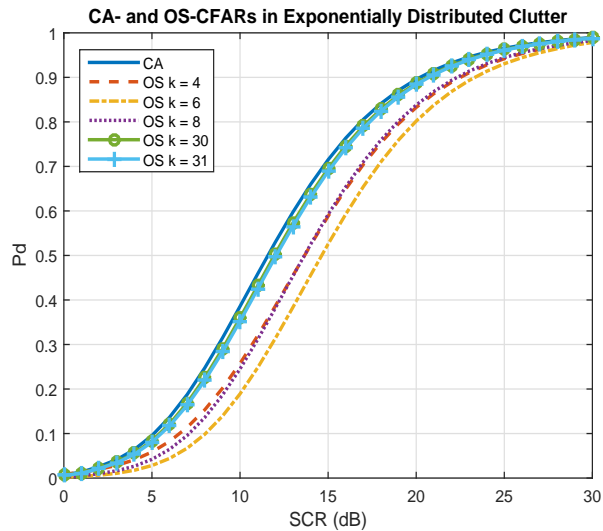}
\end{tabular}
\caption{CA- and OS-CFAR performance examples.}
\label{plot1}
\end{figure}

Figure \ref{plot1} plots the performance of the CA-CFAR together with a series of OS-CFARs. This figure shows the CA-CFAR has the best performance, while the detection performance of the OS-CFAR increases with the OS index $k$.

\begin{figure}[ht]
\centering
\begin{tabular}{c}
\includegraphics[width=0.5\textwidth,height=0.4\textwidth]{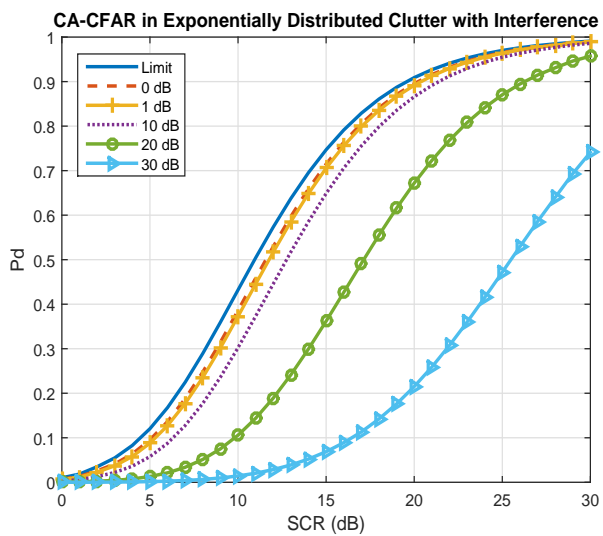}
\end{tabular}
\caption{The CA-CFAR in the presence of interference.}
\label{plot2}
\end{figure}

Next the effects of an independent Swerling I target model, inserted into the CRP, is investigated. Such an interfering target can result from a secondary target or spillover from the CUT.
Figure \ref{plot2} examines the CA-CFAR in the case of such interference. In this situation the detector \eqref{appBaux4} is included as an upper bound. The figure shows the CA-CFAR in the absence of interference (0 dB) and four cases of varying levels of interference, from 
1, 10 20 and 30 dB. The figure demonstrates that the performance of the CA-CFAR can degrade seriously in the presence of strong interference.

\begin{figure}[ht]
\centering
\begin{tabular}{c}
\includegraphics[width=0.5\textwidth,height=0.4\textwidth]{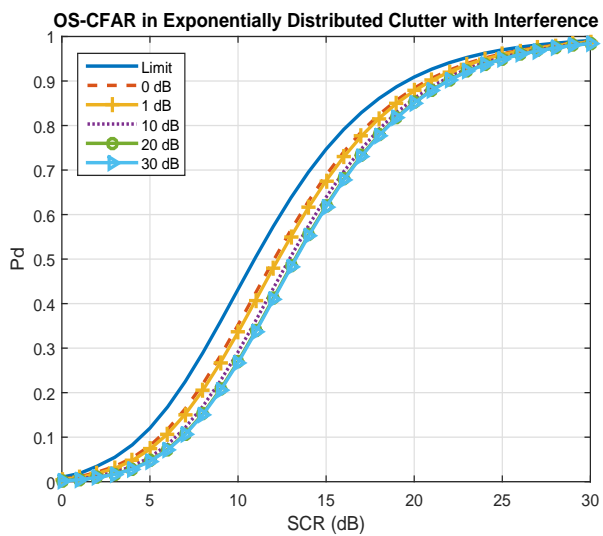}
\end{tabular}
\caption{The OS-CFAR with interference in the CRP.}
\label{plot3}
\end{figure}

Figure \ref{plot3} repeats the scenario of Figure \ref{plot2} for the OS-CFAR with $k=N-1$. Here one observes that the OS-CFAR experiences less of a detection loss in the presence of strong interference in contrast to the CA-CFAR.

\begin{figure}[ht]
\centering
\begin{tabular}{c}
\includegraphics[width=0.5\textwidth,height=0.4\textwidth]{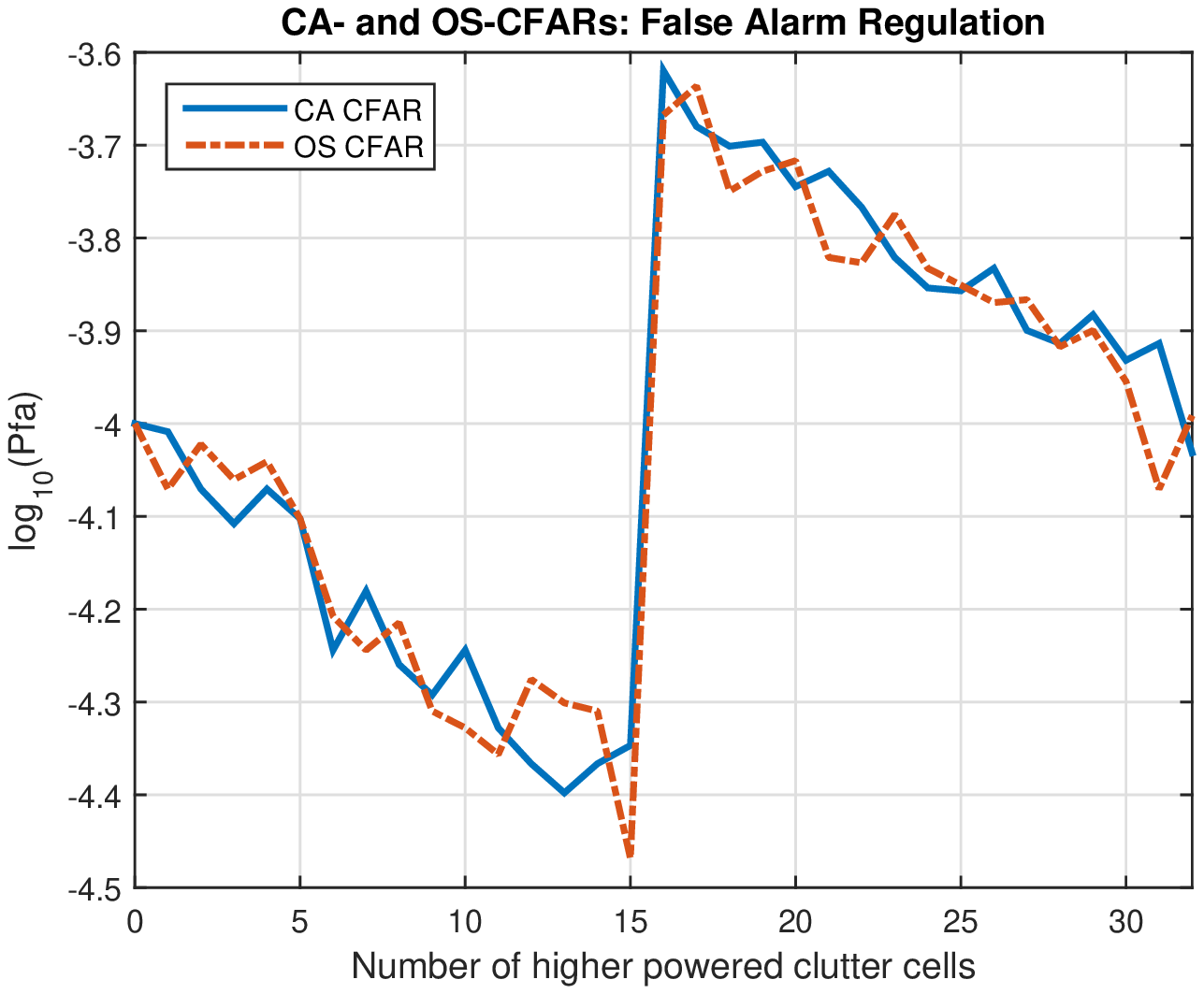}
\end{tabular}
\caption{False alarm regulation of the two CFAR detectors.}
\label{plot4}
\end{figure}

To complete the numerical analysis, the two detector's resultant Pfa is estimated when the CRP is subjected to variations in the clutter statistic's power. This is done by gradually saturating the CRP with higher powered clutter to see the effect on the design Pfa \cite{gandhi88}
Here the clutter power is measured by the mean square of the exponential distribution. In order to achieve a clutter power level increase of $x$ dB one selects an exponential distribution with parameter $\lambda \times 10^{-\frac{x}{10}}$ for the higher powered returns.
The resultant Pfa is then plotted as a function of the number of higher powered clutter returns.
Figure \ref{plot4} shows the resultant Pfa for the two CFAR detectors, where the design Pfa is $10^{-4}$ and $N = 32$ as before. One observes that as the number of higher powered clutter cells is increased, the Pfa decreases. Once the mid-point of the 
CRP is passed, the CUT is also assumed to be affected by higher power clutter, in view of Figure \ref{fig1}. This causes the characteristic jump, and thereafter the resultant Pfa exceeds the design Pfa as shown. As the full CRP is saturated the resultant Pfa limits back to the design Pfa.
This plot shows both CFAR detectors regulate the Pfa in much the same way. 

\section{Concluding Remarks}
An introduction to the theory of sliding window detectors was presented, together with a discussion of the CFAR property, and then some examples of performance of two classical detectors in exponentially distributed clutter was examined. Further developments, under a Pareto clutter model assumption, can be found in \cite{weinbergbook}.

\section*{Acknowledgements}
Comments by Dr Andrew Shaw are appreciated and improved the chapter considerably.

\end{document}